\documentclass[ a4paper, 10pt, notitlepage, twocolumn, showpacs, aps, pra, amsmath, amsfonts, amssymb ]{revtex4-1}
\RequirePackage[utf8x]{inputenc}
\RequirePackage[english]{babel}
\RequirePackage{amsmath,amsfonts,mathrsfs,amssymb,dsfont,braket,bbold}
\RequirePackage{graphicx}
\RequirePackage{color}
\definecolor{darkblue}{rgb}{0,0,.6}
\RequirePackage{hyperref} \hypersetup{colorlinks=true,linkcolor=darkblue,citecolor=darkblue,urlcolor=darkblue}

\begin{document}


\title{Pointer-based simultaneous measurements of conjugate observables in a thermal environment}
\author{Raoul Heese}\email{raoul.heese@uni-ulm.de}
\author{Matthias Freyberger}\affiliation{Institut f\"ur Quantenphysik and Center for Integrated Quantum Science and Technology \emph{(IQ\textsuperscript{ST})}, Universit\"at Ulm, D-89069 Ulm, Germany}
\date{\today}
\begin{abstract}
We combine traditional pointer-based simultaneous measurements of conjugate observables with the concept of quantum Brownian motion of multipartite systems to phenomenologically model simultaneous measurements of conjugate observables in a thermal environment. This approach provides us with a formal solution of the complete measurement dynamics for quadratic Hamiltonians and we can therefore discuss the measurement uncertainty and optimal measurement times. As a main result, we obtain a lower bound for the uncertainty of a noisy measurement, which is an extension of a previously known uncertainty relation and in which the squeezing of the system state to be measured plays an important role. This also allows us to classify minimal uncertainty states in more detail.
\end{abstract}

\maketitle


\section{Introduction} \label{sec:section 0}
The history of quantum mechanics has always been closely related to the quest for a suitable description of quantum measurements. Be it the early works \cite{born1926,heisenberg1927,schroedinger1934} or the more recent summaries \cite{braginsky1992,peres1998,clarke2014}, quantum measurements always build the framework for any further considerations. Modern measurement theories mainly concentrate on open quantum systems, decoherence, and the transition from quantum to classical \cite{schlosshauer2007}.\par
A well-known measurement theory for simultaneous measurements of conjugate observables is the theory of pointer-based measurements \cite{arthurs1965,stenholm1992}. It is based on von Neumann's idea \cite{vonneumann1932} of treating the measurement apparatus as a quantum mechanical system called \emph{pointer} from which information about a system to be measured can be inferred. So far, pointer-based simultaneous measurements have always assumed isolated quantum mechanical systems without any connection to a possible environment. We present an extension of this measurement model which can be derived from basic principles and allows exploration of a measurement configuration in a thermal environment from a phenomenological perspective. In order to achieve this goal, we combine traditional pointer-based measurements with the concept of quantum Brownian motion of multipartite systems \footnote{An overview about quantum Brownian motion of single systems can be found in, e.\,g., Ref.~\cite{hanggi2005} and references therein. Quantum Brownian motion of multipartite systems is discussed in Refs.~\cite{chou2008,anastopoulos2010,fleming2011a,fleming2011b,martinez2013}.}.\par
This approach allows us to include thermal noise into the measurement uncertainty, which is reasonable for any truly macroscopic measurement apparatus. In particular, it becomes apparent that the squeezing of the system state to be measured \cite{barnett1997} plays a crucial role in this uncertainty and simply choosing a minimal uncertainty state is no longer sufficient for an optimal measurement. Our lower bound for the uncertainty of a noisy measurement thus shows that the class of minimal uncertainty states can be understood in more detail depending on how the states react to measurement noise.\par
Moreover, there is recent work \cite{busch2013a,busch2007}, in which the authors derive a very general and state-independent uncertainty relation, which is based on the original error-disturbance concept of Heisenberg \cite{heisenberg1927}. In the context of this general framework, our model can be understood as a specific realization of a noisy measurement device, which dynamically generates disturbance.\par
To begin with, we briefly review the concept of pointer-based simultaneous measurements in Sec.~\ref{sec:review} to set the stage for Sec.~\ref{sec:operational view}, where we introduce our model from an operational point of view and solve the arising equations of motion. This preparatory work allows us to derive the uncertainty of the measurement procedure and its lower limit, an extension of a previously known uncertainty relation, in Sec.~\ref{sec:uncertainty relation}. In Sec.~\ref{sec:conclusion}, we conclude with a brief summary and outlook.


\section{A short review of pointer-based simultaneous measurements} \label{sec:review}
\begin{figure}[ht]
  \centering \includegraphics{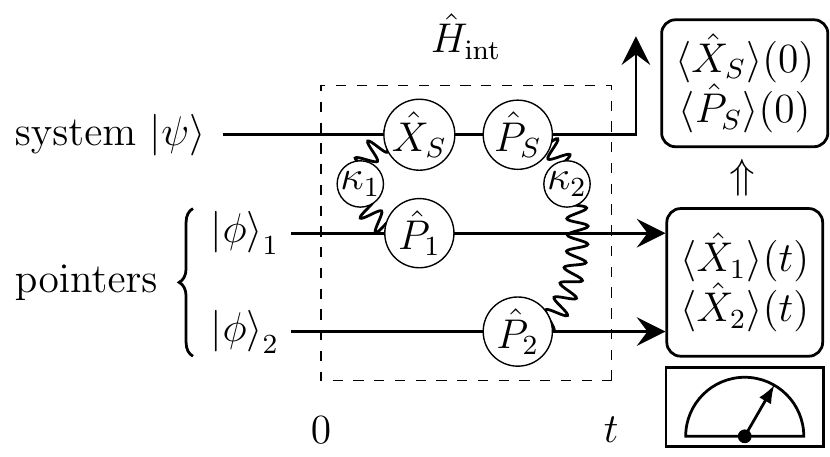}
  \caption{Schematic representation of a pointer-based measurement. The system to be measured, described by the state $\ket{\psi}$, and the pointers, described by the states $\ket{\phi}_1$ and $\ket{\phi}_2$, respectively, interact with each other by means of an interaction Hamiltonian $\hat{H}_{\mathrm{int}}$, Eq.~(\ref{eq:H:int}), which couples the position of the system to be measured, $\hat{X}_{S}$, to the first pointer's momentum $\hat{P}_{1}$ with the coupling strength $\kappa_1$ and the momentum of the system to be measured, $\hat{P}_{S}$, to the second pointer's momentum $\hat{P}_{2}$ with the coupling strength $\kappa_2$. After the interaction process at time $t$, a projective measurement of the pointer positions $\hat{X}_{1}(t)$ and $\hat{X}_{2}(t)$ is being performed. The system's initial position $\hat{X}_{S}(0)$ and momentum $\hat{P}_{S}(0)$ can then be inferred from those measurement results.}
  \label{fig:closedmodel}
\end{figure}
To introduce the conceptual idea of pointer-based simultaneous measurements, we recall a classic model \cite{vonneumann1932,arthurs1965,stenholm1992} which describes the indirect measurement of position and momentum of a free particle. Specifically, this measurement model is based on an interaction of three quantum mechanical systems with continuous variables: one system to be measured and two pointers which represent the measurement devices (see Fig.~\ref{fig:closedmodel}).\par
The system to be measured is coupled to the pointers by a bilinear interaction Hamiltonian
\begin{align} \label{eq:H:int}
  \hat{H}_{\mathrm{int}} \equiv \kappa_1 \hat{X}_S \hat{P}_1 + \kappa_2 \hat{P}_S \hat{P}_2,
\end{align}
where $\hat{X}_S$ and $\hat{P}_S$ denote the system's position and momentum, respectively. The operators $\hat{P}_1$ and $\hat{P}_2$ represent the momenta of either one of the two pointers and $\kappa_1$ and $\kappa_2$ are corresponding coupling strengths. Apart from their mutual coupling all three systems behave as free particles in this classic model.\par
Equation~(\ref{eq:H:int}) is in fact just a sum of the exponents of two displacement operators and therefore the dynamical behavior of the model is straightforward: the position of the system to be measured displaces the position of the first pointer and its corresponding momentum displaces the position of the second pointer. As a consequence, projectively measuring the pointer positions after the coupling interaction allows one to infer the position and the momentum of the system to be measured. Since both pointers are assumed to be independent systems, they can be measured independently. The accuracy of this measurement is influenced by the measurement devices and the coupling strengths and can for example be quantified by variances \cite{arthurs1965,she1966,wodkiewicz1987,arthurs1988,stenholm1992,ozawa2004,busshardt2010,busshardt2011} or information entropy \cite{buzek1995,heese2013}.\par
However, Eq.~(\ref{eq:H:int}) represents just one possible choice of interaction for realizing a pointer-based measurement. An alternative coupling could also be chosen in such a way that the system displaces the momenta of the pointers instead of their positions and allows measurement of these momenta to infer the system position and momentum. Moreover, different noncommuting continuous variables like, for example, quadratures \cite{scully1997,schleich2001} of the quantized electromagnetic field, could be used instead of position and momentum. The coupling strengths between system and pointers can also be time dependent \cite{busshardt2010,busshardt2011}.\par
Nevertheless, the model outlined here describes only the measurement of a system with measurement devices totally isolated from their environment. However, for a more complete and more realistic treatment of a measurement process it is necessary to include environmental effects. In other words, each pointer has to experience its environment since it represents a macroscopic apparatus. The corresponding environmental effects are naturally expected to cause noise on the measurement result, but furthermore may also introduce decoherence \cite{zurek2003} and are possibly a first step in eliminating the need for a final projective ``ideal single variable measurement" \cite{she1966} of the pointer observables, which is a crucial point of criticism in any theory of quantum measurements \footnote{It has been pointed out that the concept of decoherence on its own does not completely solve the measurement problem. See \cite{leggett2002,adler2003,zurek2003}.}. In the following sections, we present an approach towards a theory for such open pointer-based measurements as a generalization of closed pointer-based measurements.


\section{An operational view on open pointer-based simultaneous measurements} \label{sec:operational view}
There are several ways to model an environment, e.\,g., specialized quantization procedures \cite{bolivar1998}, the concept of stochastic Schr{\"o}dinger equations \cite{bassi2013} or system-plus-reservoir approaches \cite{ford1988}, which are exhaustively described in, e.\,g., Refs.~\cite{carmichael1991,weiss1999} and references therein. Inspired by Refs.~\cite{fleming2011a,fleming2011b}, it turned out that a system-plus-reservoir approach, implemented by a bilinearly coupled bosonic heat bath of a collection of harmonic oscillators, is most suitable for our purposes. Briefly put, we utilize the well-known Caldeira-Leggett model \cite{caldeira198183}.\par
In this section, we first present a model for open pointer-based simultaneous measurements and then introduce a suitable rescaling, which allows us to express the equations of motion in dimensionless quantities. We subsequently argue that a renormalization of these equations of motion is necessary to eliminate unphysical terms. Finally, we present a solution of the renormalized equations of motion.

\subsection{Model}
In accordance with the closed pointer-based simultaneous measurement in Sec.~\ref{sec:review}, we consider two pointer particles of identical mass $M$, which are coupled to the system particle to be measured of mass $M_S$ via the classic interaction Hamiltonian, Eq.~(\ref{eq:H:int}). Their positions and momenta read $\hat{\mathbf{X}}^{T} \equiv (\hat{X}_S, \hat{X}_1, \hat{X}_2)$ and $\hat{\mathbf{P}}^{T} \equiv (\hat{P}_S, \hat{P}_1, \hat{P}_2)$, respectively. As already mentioned above, our environment consists of $N$ harmonic oscillators, i.\,e., $N$ particles of mass $m$ with positions $\hat{\mathbf{q}}^{T} \equiv (\hat{q}_1, \dots, \hat{q}_N)$ and momenta $\hat{\mathbf{k}}^{T} \equiv (\hat{k}_1, \dots, \hat{k}_N)$, which are bilinearly coupled to each other by means of the bath Hamiltonian
\begin{align} \label{eq:H:bath}
  \hat{H}_{\mathrm{bath}} \equiv \frac{1}{2 m} \hat{\mathbf{k}}{}^{T}\hat{\mathbf{k}} + \frac{1}{2} \hat{\mathbf{q}}{}^{T} \mathbf{c} \hat{\mathbf{q}}
\end{align} 
with the bath-internal coupling matrix $\mathbf{c}$.\par
By definition, system and pointers couple to the environment bilinearly via their positions, so we introduce an environmental Hamiltonian
\begin{align} \label{eq:H:env}
  \hat{H}_{\mathrm{env}} \equiv \hat{\mathbf{q}}^{T} \mathbf{g} \hat{\mathbf{X}}
\end{align} 
with the environmental coupling matrix $\mathbf{g}$. Although the form of Eq.~(\ref{eq:H:env}) is typical for system-plus-reservoir approaches, it would also be a valid assumption to let system and pointers couple to the environment via their momenta \cite{ford1988}. Yet we do not further pursue these considerations here, in favor of a simple model.\par
So far, the coupling matrices $\mathbf{c}$ and $\mathbf{g}$ can be chosen completely freely and determine the structure of the bath and its influence on both system and pointers. However, for reasons of clarity, we do not specify them before we perform the transition to continuous bath modes in Sec.~\ref{sec:operational view:equations of motion}.\par
Summarized, the complete Hamiltonian $\hat{\mathscr{H}}$ of our model for open pointer-based simultaneous measurements consists of the free-particle Hamiltonian of system and pointers
\begin{align} \label{eq:H:free}
  \hat{H}_{\mathrm{free}} \equiv \frac{\hat{P}_S^2}{2 M_S} + \frac{\hat{P}_1^2}{2 M} + \frac{\hat{P}_2^2}{2 M},
\end{align}
the classic interaction Hamiltonian $\hat{H}_{\mathrm{int}}$, Eq.~(\ref{eq:H:int}), the bath Hamiltonian $\hat{H}_{\mathrm{bath}}$, Eq.~(\ref{eq:H:bath}), and the environmental Hamiltonian $\hat{H}_{\mathrm{env}}$, Eq.~(\ref{eq:H:env}), and therefore reads
\begin{align} \label{eq:H}
  \hat{\mathscr{H}} \equiv \hat{H}_{\mathrm{free}} + \hat{H}_{\mathrm{int}} + \hat{H}_{\mathrm{bath}} + \hat{H}_{\mathrm{env}}.
\end{align}\par
To understand the dynamics of our model, our main interest lies in the system and pointer positions $\hat{\mathbf{X}}(t)$ and momenta $\hat{\mathbf{P}}(t)$, respectively, in the Heisenberg picture. Their operational behavior sets the framework for our discussion in Sec.~\ref{sec:uncertainty relation}, where we mainly deal with their first and second moments in order to infer measurement results with an associated uncertainty. Therefore, we also need knowledge about the initial quantum mechanical state of the measurement setup.\par
It is physically reasonable to assume that the system to be measured, represented by the state $\ket{\psi}$, and the two pointers, represented by the states $\ket{\phi}_1$ and $\ket{\phi}_2$, respectively, are initially uncorrelated; cf. Fig.~\ref{fig:closedmodel}. Apart from this limitation, $\ket{\psi}$, $\ket{\phi}_1$, and $\ket{\phi}_2$ can be chosen completely freely. In addition, we assume that our bath is initially in a thermal state
\begin{align}\label{eq:thermalstate}
  \hat{\varrho}_{\mathrm{bath}}(0) \equiv \frac{1}{Z} \exp \left[ - \beta \hat{H}_{\mathrm{bath}} \right]
\end{align}
of thermal energy $\beta^{-1} \equiv k_B \vartheta$ with the Boltzmann constant $k_B$ and the temperature $\vartheta$. Here we use the bath Hamiltonian $\hat{H}_{\mathrm{bath}}$, Eq.~(\ref{eq:H:bath}), and the normalizing partition function $Z$. As a consequence, the bath is also initially uncorrelated with the system or the pointers \footnote{For a discussion of baths initially correlated with the system or pointers or distinct baths of different temperature, we refer to Ref.~\cite{grabert1988,fleming2011c,fleming2012} and references therein. Such baths might change the structure of our results in Sec.~\ref{sec:uncertainty relation}, e.\,g., by introducing correlation terms.}.\par
In short, the initial state $\hat{\varrho}(0)$ of the measurement configuration can be summarized as
\begin{align} \label{eq:initialstate}
  \hat{\varrho}(0) \equiv \ket{\psi} \bra{\psi} \otimes \ket{\phi}_1 {\vphantom{\ket{\phi}}}_1\!\bra{\phi} \otimes \ket{\phi}_2 {\vphantom{\ket{\phi}}}_2\!\bra{\phi} \otimes \hat{\varrho}_{\mathrm{bath}}(0).
\end{align}
The complete Hamiltonian, Eq.~(\ref{eq:H}), and the complete initial state, Eq.~(\ref{eq:initialstate}), set the framework for all further considerations.

\subsection{Rescaling}
In order to reduce our equations to their most fundamental ingredients and to eliminate all units from our variables, we pursue a rescaling approach similar to that of Ref.~\cite{busshardt2010}. For this purpose, we identify a characteristic time scale
\begin{align} \label{eq:T}
  T \equiv \frac{\Delta X_S(0) M_S}{\Delta P_S(0)},
\end{align}
with the initial variances of system position
\begin{subequations} \label{eq:systemvariance}
\begin{align} \label{eq:systemvariance:x}
  \Delta X_S^2(0)  \equiv \braket{\hat{X}_S(0)^2} - \braket{\hat{X}_S(0)}^2
\end{align}
and momentum
\begin{align} \label{eq:systemvariance:p}
  \Delta P_S^2(0)  \equiv \braket{\hat{P}_S(0)^2} - \braket{\hat{P}_S(0)}^2,
\end{align}
\end{subequations}
respectively, which corresponds to the typical spreading time scale of the system's initial wave function if the system were isolated from the pointers and the bath. In other words, $T$ describes the time scale during which a free test particle can be considered localized inside our measurement apparatus. We define our rescaled time $t'$ in units of this measurement time scale, so that $t' \equiv t/T$. The associated energy $\hbar/T$ stands for the interaction energy scale of this measurement process and leads us to the rescaled Hamiltonian $\mathscr{H}' \equiv \mathscr{H} T / \hbar$, Eq.~(\ref{eq:H}). We deal with the rescaled thermal energy of the bath $\beta'{}^{-1} \equiv \beta^{-1} T / \hbar$, Eq.~(\ref{eq:thermalstate}), in the same way.\par
Furthermore, we make use of a corresponding characteristic length
\begin{align}
  \lambda \equiv \sqrt{\frac{T \hbar}{M_S}},
\end{align}
which represents the typical spreading length scale of the system's initial wave function if it were isolated from the pointers and the bath in the same way as for the characteristic time scale $T$, Eq.~(\ref{eq:T}). Specifically, we rewrite the system, pointer, and bath coordinates as $\hat{\mathbf{X}}' \equiv \hat{\mathbf{X}}/\lambda$, $\hat{\mathbf{P}}' \equiv \hat{\mathbf{P}} \lambda / \hbar$, $\hat{\mathbf{q}}' \equiv \hat{\mathbf{q}}/\lambda$, and $\hat{\mathbf{k}}' \equiv \hat{\mathbf{k}} \lambda / \hbar$ and define the rescaled bath-internal coupling matrix $\mathbf{c}$, Eq.~(\ref{eq:H:bath}), as $\mathbf{c}' \equiv \mathbf{c} T^2/M_S$, the rescaled position-based coupling to the environment $\mathbf{g}$, Eq.~(\ref{eq:H:env}), as $\mathbf{g}' \equiv \mathbf{g} T^2/M_S$, and the rescaled system-pointer coupling strengths $\kappa_1$ and $\kappa_2$, Eq.~(\ref{eq:H:int}), as $\kappa_1' \equiv \kappa_1 T M / M_S$ and $\kappa_2' \equiv \kappa_2 M$, respectively.\par
As a consequence, the rescaled Hamiltonian reads
\begin{align} \label{eq:H:rescaled}
  \hat{\mathscr{H}}' = & \phantom{+.} \frac{\hat{P}_S'{}^2}{2} + \frac{\hat{P}_1'{}^2}{2 M_0} + \frac{\hat{P}_2'{}^2}{2 M_0} + \frac{\kappa_1'}{M_0} \hat{X}_S' \hat{P}_1' + \frac{\kappa_2'}{M_0} \hat{P}_S' \hat{P}_2' \nonumber \\
  & + \frac{1}{2 m_0} \hat{\mathbf{k}}'{}^{T} \hat{\mathbf{k}}' + \frac{1}{2} \hat{\mathbf{q}}'{}^{T} \mathbf{c}' \hat{\mathbf{q}}' + \hat{\mathbf{q}}'{}^{T} \mathbf{g}' \hat{\mathbf{X}}'.
\end{align}
with the pointer-system mass ratio
\begin{align} \label{eq:M0}
  M_0 \equiv \frac{M}{M_S}
\end{align}
and the bath-system mass ratio
\begin{align}
  m_0 \equiv \frac{m}{M_S}.
\end{align}
In particular, Eq.~(\ref{eq:H:rescaled}) leads to the rescaled Heisenberg equations
\begin{align} \label{eq:Heisenberg}
\frac{\partial}{\partial t'} \hat{A}'(t') = i \left[ \hat{\mathscr{H}}', \hat{A}'(t') \right]
\end{align}
for any rescaled observable $\hat{A}'(t')$. In the following, our aim is to solve Eq.~(\ref{eq:Heisenberg}) for both $\hat{\mathbf{X}}'(t')$ and $\hat{\mathbf{P}}'(t')$. So far, all rescaled quantities have been marked with a prime. To simplify the further notation, we drop this prime and limit ourselves exclusively to rescaled variables.

\subsection{Equations of motion} \label{sec:operational view:equations of motion}
In the spirit of Ref.~\cite{ford1988}, we first solve the Heisenberg equation for the bath oscillator observables $\hat{\mathbf{q}}^{T}(t) \equiv (\hat{q}_1(t), \dots, \hat{q}_N(t))$ and $\hat{\mathbf{k}}^{T}(t) \equiv (\hat{k}_1(t), \dots, \hat{k}_N(t))$, Eq.~(\ref{eq:Heisenberg}), and then use the results to rewrite the Heisenberg equations for the system and pointer observables $\hat{\mathbf{X}}(t)$ and $\hat{\mathbf{P}}(t)$, Eq.~(\ref{eq:Heisenberg}), as systems of generalized Langevin equations \cite{burton1983,coffey1996,hanggi2005}
\begin{subequations} \label{eq:equationsofmotion}
\begin{align} \label{eq:X:discrete}
  & \begin{pmatrix} - a \ddot{\hat{X}}_S(t) \\ M_0 \ddot{\hat{X}}_1(t) \\ - a M_0 \ddot{\hat{X}}_2(t) \end{pmatrix} + a \kappa_2 \begin{pmatrix} \ddot{\hat{X}}_2(t) \\ 0 \\ \ddot{\hat{X}}_S(t) \end{pmatrix} + \kappa_1 \begin{pmatrix} \dot{\hat{X}}_1(t) \\ - \dot{\hat{X}}_S(t) \\ 0 \end{pmatrix} \nonumber \\
  & - \frac{\kappa_1^2}{M_0} \begin{pmatrix} \hat{X}_S(t) \\ 0 \\ 0 \end{pmatrix} - \int \limits_{0}^{t} \! \mathrm{d}s \boldsymbol{\mu}(t-s) \begin{pmatrix} \hat{X}_S(s) \\ \hat{X}_1(s) \\ \hat{X}_2(s) \end{pmatrix} \nonumber \\
  = & \phantom{-.} \hat{\boldsymbol{\xi}}(t)
\end{align}
with momenta
\begin{align} \label{eq:P}
  \begin{pmatrix} \hat{P}_S(t) \\ \hat{P}_1(t) \\ \hat{P}_2(t) \end{pmatrix} = \begin{pmatrix} - a \dot{\hat{X}}_S(t) + a \kappa_2 \dot{\hat{X}}_2(t) \\ M_0 \dot{\hat{X}}_1(t) - \kappa_1 \hat{X}_S(t) \\ - a M_0 \dot{\hat{X}}_2(t) + a \kappa_2 \dot{\hat{X}}_S(t) \end{pmatrix}
\end{align}
\end{subequations}
and the abbreviation 
\begin{align} \label{eq:a}
  a \equiv \frac{M_0}{\kappa_2^2-M_0}.
\end{align}
To keep our notation simple, we limit ourselves to interaction times $t\geq0$ and consider all time-dependent quantities as confined to this regime.\par
The environmental influence on the equations of motion for system and pointers is therefore governed by two expressions, namely, the dissipation kernel \cite{fleming2011b}
\begin{align} \label{eq:dissipationkernel:discrete}
  \boldsymbol{\mu}(t) \equiv \frac{1}{m_0} \mathbf{g}^{T} \frac{\sin ( \boldsymbol{\omega} t )}{\boldsymbol{\omega}} \mathbf{g},
\end{align}
which results from the retarded propagator of the inhomogeneous bath dynamics \cite{ford1988} and determines the damping influence of the environment, and the stochastic force \cite{weiss1999}
\begin{align} \label{eq:stochasticforce}
  \hat{\boldsymbol{\xi}}(t) \equiv - \mathbf{g}^{T} \Big[ & \cos (\boldsymbol{\omega} t) \hat{\mathbf{q}}(0) + \frac{\sin (\boldsymbol{\omega} t)}{m_0 \boldsymbol{\omega}} \hat{\mathbf{k}}(0) \Big],
\end{align}
which results from the free bath dynamics and determines the noisy influence of the environment. Since the coupling matrix $\mathbf{c}$ is by definition symmetric and real, the bath frequency matrix \cite{fleming2011b}
\begin{align} \label{eq:omega}
  \boldsymbol{\omega} \equiv \sqrt{\frac{\mathbf{c}}{m_0}}
\end{align}
can always be diagonalized to describe the bath dynamics in terms of normal modes.\par
We remark that the first moment of the stochastic force, Eq.~(\ref{eq:stochasticforce}), obeys
\begin{align} \label{eq:estochasticforce}
 \braket{\hat{\boldsymbol{\xi}}(t)} = 0.
\end{align}
Moreover, the well-known fluctuation-dissipation theorem \cite{weiss1999} connects the dissipation kernel, Eq.~(\ref{eq:dissipationkernel:discrete}), with the symmetric autocorrelation function \cite{fleming2011b}
\begin{align} \label{eq:autocorrelation:discrete}
  \boldsymbol{\nu}(t_1-t_2) & \equiv \frac{1}{2} \braket{ \hat{\boldsymbol{\xi}}(t_1) \hat{\boldsymbol{\xi}}{}^{T}(t_2) + \hat{\boldsymbol{\xi}}(t_2) \hat{\boldsymbol{\xi}}{}^{T}(t_1) } \nonumber \\
  & = \frac{1}{2 m_0} \mathbf{g}^{T} \coth \left( \frac{\beta \boldsymbol{\omega}}{2} \right) \frac{\cos [ \boldsymbol{\omega} (t_1-t_2) ]}{\boldsymbol{\omega}} \mathbf{g}
\end{align}
of the stochastic force.\par
It is a common approach to switch from a bath of $N$ discrete oscillators to a continuous bath with $N \rightarrow \infty$. In this limit, Eqs.~(\ref{eq:dissipationkernel:discrete}) and (\ref{eq:autocorrelation:discrete}) can be written as integrals over the bath frequencies $\omega$, so that
\begin{align} \label{eq:dissipationkernel:I}
  \boldsymbol{\mu}(t) = \int \limits_{0}^{\infty} \! \mathrm{d} \omega \sin ( \omega t ) \mathbf{I}(\omega)
\end{align}
and
\begin{align} \label{eq:autocorrelation:I}
  \boldsymbol{\nu}(t) = \frac{1}{2} \int \limits_{0}^{\infty} \! \mathrm{d} \omega \coth \left( \frac{\beta \omega}{2} \right) \cos ( \omega t ) \mathbf{I}(\omega)
\end{align}
with the spectral density
\begin{align} \label{eq:spectraldensity}
  \mathbf{I}(\omega) \equiv \frac{1}{m_0} \mathbf{g}^{T} \omega^{-1} \delta( \omega \mathds{1} - \boldsymbol{\omega} ) \mathbf{g}.
\end{align}
Here $\delta( \omega \mathds{1} - \boldsymbol{\omega} )$ represents the Dirac delta distribution and $\mathds{1}$ stands for the identity matrix. Since the yet undetermined coupling matrix $\mathbf{c}$, Eq.~(\ref{eq:H:bath}), which determines $\boldsymbol{\omega}$, Eq.~(\ref{eq:omega}), and the also so far undetermined coupling matrix $\mathbf{g}$, Eq.~(\ref{eq:H:env}), have infinite degrees of freedom in a continuous bath, we can in principle specify them so as to design the spectral density $\mathbf{I}(\omega)$, Eq.~(\ref{eq:spectraldensity}), as a smooth function of $\omega$. This function then describes how the bath modes affect system and pointers without the need for a detailed consideration of the bath properties. Consequently, taking the continuum limit allows us to change our point of view from a microscopic to a phenomenological perspective.\par
For our model of open pointer-based simultaneous measurements, we choose two identical but separate Ohmic baths with an algebraic cutoff with cutoff frequency $\omega_c$ and frequency-independent viscosity $\eta$ \cite{weiss1999}, which are each equally coupled to either one of the pointers, whereas the system to be measured itself is not directly influenced by the environment, i.\,e., we use
\begin{align} \label{eq:spectraldensity:ohmic}
  \mathbf{I}(\omega) \equiv \frac{2 \eta}{\pi} \frac{\omega}{\frac{\omega^2}{\omega_c^2} +1} \begin{pmatrix} 0 & 0 & 0 \\ 0 & 1 & 0 \\ 0 & 0 & 1 \end{pmatrix}.
\end{align}
As a consequence, Eqs.~(\ref{eq:dissipationkernel:I}) and (\ref{eq:autocorrelation:I}) can be written as
\begin{align} \label{eq:dissipationkernel}
  \boldsymbol{\mu}(t) = \eta \omega_c^2 e^{-\omega_c t} \begin{pmatrix} 0 & 0 & 0 \\ 0 & 1 & 0 \\ 0 & 0 & 1 \end{pmatrix}
\end{align}
and
\begin{align} \label{eq:autocorrelation}
  \boldsymbol{\nu}(t) = & \int \limits_{0}^{\infty} \! \mathrm{d} \omega \frac{\omega \coth \left( \frac{\beta \omega}{2} \right) \cos ( \omega t )}{\omega^2+\omega_c^2} \frac{\eta \omega_c^2}{\pi} \begin{pmatrix} 0 & 0 & 0 \\ 0 & 1 & 0 \\ 0 & 0 & 1 \end{pmatrix},
\end{align}
respectively. Thus, the phenomenological equations of motion for our system and pointer observables are given by Eq.~(\ref{eq:equationsofmotion}) with the dissipation kernel chosen in Eq.~(\ref{eq:dissipationkernel}) and a stochastic force, Eq.~(\ref{eq:stochasticforce}), which obeys Eqs.~(\ref{eq:estochasticforce}) and (\ref{eq:autocorrelation}).

\subsection{Renormalization} \label{sec:operational view:renormalization}
System-plus-reservoir models have the tendency to bear subtle but well-known complications, which have to be treated with special care; see, e.\,g., Refs.~\cite{weiss1999,hanggi2005,fleming2011a} and references therein for a detailed discussion. The reasons for this are mainly the unphysical starting conditions of our model, more specifically the sudden interaction of the initially uncorrelated bath with the system and pointer states. From a phenomenological point of view, it is therefore reasonable to adjust our original model in such a way that its equations of motion, Eq.~(\ref{eq:equationsofmotion}), are free from any unphysical artifacts. There are various attempts to achieve this goal (see, e.\,g., Refs.~\cite{weiss1999,hanggi2005,fleming2011a} and references therein), and we present only one straightforward strategy here without discussing all of the possible alternatives.\par
Specifically, our equations of motion include two obvious unphysical artifacts, namely, the so-called potential shift and the so-called slip term, both of which are contained within the second and third components of the integral term in Eq.~(\ref{eq:X:discrete}). By inserting Eq.~(\ref{eq:dissipationkernel}) into Eq.~(\ref{eq:X:discrete}), we can express these integral components as
\begin{align} \label{eq:integralterm}
  & - \eta \omega_c^2 \int \limits_{0}^{t} \! \mathrm{d}s e^{- (t-s) \omega_c} \begin{pmatrix} \hat{X}_1(s) \\ \hat{X}_2(s) \end{pmatrix} \nonumber \\
  = & - \eta \omega_c \begin{pmatrix} \hat{X}_1(t) \\ \hat{X}_2(t) \end{pmatrix} + \eta \omega_c e^{- t \omega_c} \begin{pmatrix} \hat{X}_1(0) \\ \hat{X}_2(0) \end{pmatrix} \nonumber \\
  & + \eta \omega_c \int \limits_{0}^{t} \! \mathrm{d}s e^{- (t-s) \omega_c} \begin{pmatrix} \dot{\hat{X}}_1(s) \\ \dot{\hat{X}}_2(s) \end{pmatrix}
\end{align}
to identify the two crucial expressions: The first term on the right-hand side of Eq.~(\ref{eq:integralterm}) represents the potential shift. This term has the same influence on our equations of motion as an external potential which is quadratic in position and proportional to the cutoff frequency $\omega_c$. The potential shift is therefore cutoff sensitive and even becomes divergent in the limit $\omega_c \rightarrow \infty$. As a consequence, it must be considered unphysical.\par
The second term on the right-hand side of Eq.~(\ref{eq:integralterm}) is the slip term, which effectively represents an additional external force with a strength proportional to the cutoff frequency $\omega_c$ occurring on the time scale $\omega_c^{-1}$. Since this external force leads to an effective stochastic force with different properties from those of our original stochastic force, Eq.~(\ref{eq:stochasticforce}), we consider it an unphysical artifact. In the high-cutoff limit 
\begin{align} \label{eq:highcutofflimit}
   \omega_c \tau \gg 1,
\end{align}
where $\tau$ represents the typical measurement time scale of our measurement device, i.\,e., the largest relevant interaction time, we can approximately write
\begin{align} \label{eq:highcutoff}
   \omega_c e^{- t \omega_c} & \approx 2 \delta(t)
\end{align}
with the Dirac delta distribution $\delta(t)$. In other words, in the high-cutoff limit the slip term becomes a mere initial kick.\par
Thus, the two unphysical terms on the right-hand side of Eq.~(\ref{eq:integralterm}) can be written as 
\begin{align} \label{eq:integralterm:highcutoff}
   & - \eta \omega_c \begin{pmatrix} \hat{X}_1(t) \\ \hat{X}_2(t) \end{pmatrix} + \eta \omega_c e^{- t \omega_c} \begin{pmatrix} \hat{X}_1(0) \\ \hat{X}_2(0) \end{pmatrix} \nonumber \\
  \approx  & \phantom{-} \eta ( 2 \delta(t) - \omega_c ) \begin{pmatrix} \hat{X}_1(t) \\ \hat{X}_2(t) \end{pmatrix}
\end{align}
in the regime given by Eq.~(\ref{eq:highcutofflimit}). By performing the renormalization \cite{fleming2011a}
\begin{align} \label{eq:renormalization1}
  \mathscr{H} \longrightarrow \mathscr{H} + \eta \left[ \frac{\omega_c}{2} - \delta(t) \right] ( \hat{X}_1^2 + \hat{X}_2^2 )
\end{align}
of the Hamiltonian $\mathscr{H}$, Eq.~(\ref{eq:H:rescaled}), we can straightforwardly eliminate Eq.~(\ref{eq:integralterm:highcutoff}) from our equations of motion. This behavior is also related to a translation-symmetric environmental coupling \cite{ingold2002}.\par
As a consequence, only the third term on the right-hand side of Eq.~(\ref{eq:integralterm}) remains and thus Eq.~(\ref{eq:X:discrete}) reads
\begin{align} \label{eq:X}
  & \begin{pmatrix} - a \ddot{\hat{X}}_S(t) \\ M_0 \ddot{\hat{X}}_1(t) \\ - a M_0 \ddot{\hat{X}}_2(t) \end{pmatrix} + a \kappa_2 \begin{pmatrix} \ddot{\hat{X}}_2(t) \\ 0 \\ \ddot{\hat{X}}_S(t) \end{pmatrix} + \kappa_1 \begin{pmatrix} \dot{\hat{X}}_1(t) \\ - \dot{\hat{X}}_S(t) \\ 0 \end{pmatrix} \nonumber \\
  & - \frac{\kappa_1^2}{M_0} \begin{pmatrix} \hat{X}_S(t) \\ 0 \\ 0 \end{pmatrix} + \eta \omega_c \int \limits_{0}^{t} \! \mathrm{d}s e^{- (t-s) \omega_c} \begin{pmatrix} 0 \\ \dot{\hat{X}}_1(s) \\ \dot{\hat{X}}_2(s) \end{pmatrix} \nonumber \\
  = & \phantom{-.} \hat{\boldsymbol{\xi}}(t)
\end{align}
while Eq.~(\ref{eq:P}) remains unchanged. In the following, we use these renormalized equations of motion.\par
Note, however, that additional unphysical artifacts may also occur which are more difficult to spot and cannot be eliminated by a simple renormalization \cite{fleming2011a,fleming2011d}, like a sudden initial jolt in physical quantities on a time scale $\omega_c^{-1}$ \cite{hu1992} or a ``spurious" logarithmic cutoff-sensitivity of system correlators \cite{hu2004,fleming2011d}. Some of these effects could be eliminated by a smooth switch-on of the interaction with the bath \cite{fleming2011a}. We believe that such an approach leads to more complicated expressions but does not change our final results. Since we also do not expect that additional artifacts play an important role in our phenomenological model, we do not apply further strategies to suppress them.

\subsection{Formal solution} \label{sec:operational view:formal solution}
It is well known that by transforming to Laplace space, the renormalized equations of motion, Eqs.~(\ref{eq:X}) and (\ref{eq:P}), become purely algebraic equations which can then be solved and transformed back to the time domain. As a result, we get
\begin{subequations} \label{eq:solutions}
\begin{align} \label{eq:solutions:X}
  \hat{\mathbf{X}}(t) = \mathbf{K}(t) \hat{\mathbf{X}}(0) + \mathbf{G}(t) \hat{\mathbf{P}}(0) + \hat{\mathbf{\Lambda}}(t)
\end{align}
and
\begin{align} \label{eq:solutions:P}
  \hat{\mathbf{P}}(t) = & \phantom{+.} \left[ \mathbf{M} \dot{\mathbf{K}}(t) - \mathbf{D} \mathbf{K}(t) \right] \hat{\mathbf{X}}(0) \nonumber \\
  & + \left[ \mathbf{M} \dot{\mathbf{G}}(t) - \mathbf{D} \mathbf{G}(t) \right] \hat{\mathbf{P}}(0) \nonumber \\
  & + \mathbf{M} \dot{\hat{\mathbf{\Lambda}}}(t) - \mathbf{D} \hat{\mathbf{\Lambda}}(t),
\end{align}
\end{subequations}
respectively. Here, we have introduced the noise
\begin{align} \label{eq:noise}
  \hat{\mathbf{\Lambda}}(t) & \equiv \begin{pmatrix} \hat{\Lambda}_S(t) \\ \hat{\Lambda}_1(t) \\ \hat{\Lambda}_2(t) \end{pmatrix} \equiv \int \limits_{0}^{t} \! \mathrm d s\ \mathbf{G}(t-s) \hat{\boldsymbol{\xi}}(s),
\end{align}
and the propagators
\begin{subequations} \label{eq:propagators}
\begin{align} \label{eq:propagator:G}
  \mathbf{G}(t) \equiv \mathcal{L}^{-1} \left\{ \begin{pmatrix} -u(s) & \kappa_1 s & a \kappa_2 s^2 \\ - \kappa_1 s & b(1,s) & 0 \\ a \kappa_2 s^2 & 0 & b(-a,s) \end{pmatrix}^{-1} \right\}(t)
\end{align}
and
\begin{align} \label{eq:propagator:K}
  \mathbf{K}(t) \equiv \dot{\mathbf{G}}(t) \mathbf{M} + \mathbf{G}(t) \mathbf{D}^{T},
\end{align}
\end{subequations}
respectively, where $\mathcal{L}^{-1}\{f(s)\}(t)$ denotes the inverse Laplace transform of a function $f(s)$ in the frequency domain. The abbreviations $a$, Eq.~(\ref{eq:a}), and $u(s) \equiv a s^2 + M_0^{-1} \kappa_1^2$ simplify our notation, and the function
\begin{align}
  b(w,s) & \equiv \frac{w s}{s+\omega_c} \left( M_0 s^2 + M_0 s \omega_c + \frac{\eta \omega_c}{w} \right)
\end{align}
represents the bath-dependent parts of the propagators. Furthermore, we make use of the coupling matrices
\begin{subequations}
\begin{align} 
  \mathbf{M} \equiv \begin{pmatrix} -a & 0 & a \kappa_2 \\ 0 & M_0 & 0 \\ a \kappa_2 & 0 & - a M_0 \end{pmatrix}
\end{align}
and
\begin{align} 
  \mathbf{D} \equiv \begin{pmatrix} 0 & 0 & 0 \\ \kappa_1 & 0 & 0 \\ 0 & 0 & 0 \end{pmatrix},
\end{align}
\end{subequations}
respectively.\par
Explicitly performing the inverse Laplace transform in Eq.~(\ref{eq:propagator:G}) technically corresponds to finding the nontrivial zeros of polynomials of up to sixth order. Therefore, we limit ourselves to a formal notation. Note that by turning off the influence of the bath (i.\,e., $\eta = 0$), Eq.~(\ref{eq:solutions}) represents the solution of the closed pointer-based measurement \cite{busshardt2010}. Furthermore note that Eq.~(\ref{eq:propagator:G}) is defined only for $1/a \neq 0$, i.\,e., $\kappa_2 \neq \pm \sqrt{M_0}$, Eq.~(\ref{eq:a}), which is the condition under which a nonsingular Lagrangian \cite{cisneros-parra2012} exists for our model.\par
In brief, the dynamics of the system and pointer observables $\hat{\mathbf{X}}(t)$, Eq.~(\ref{eq:solutions:X}), and $\hat{\mathbf{P}}(t)$, Eq.~(\ref{eq:solutions:P}), respectively, can be expressed in terms of propagators acting on their initial values under the noisy influence of the bath, which forces them into a quantum Brownian motion. However, for a pointer-based measurement as described in Sec.~\ref{sec:review}, the dynamics of the system to be measured and the pointer momentum dynamics are not of central interest and it is therefore sufficient to concentrate on the pointer position dynamics $\hat{X}_1(t)$ and $\hat{X}_2(t)$.

\subsection{Pointer position dynamics}
In order to get the pure pointer position dynamics, Eq.~(\ref{eq:solutions:X}) can be reduced to
\begin{align} \label{eq:solutions:PX}
  \begin{pmatrix} \hat{X}_1(t) \\ \hat{X}_2(t) \end{pmatrix} = \mathbf{A}(t) \begin{pmatrix} \hat{X}_S(0) \\ \hat{P}_S(0) \end{pmatrix} + \mathbf{B}(t) \hat{\mathbf{J}} + \begin{pmatrix} \hat{\Lambda}_1(t) \\ \hat{\Lambda}_2(t) \end{pmatrix}
\end{align}
with the coefficients
\begin{subequations} \label{eq:inferredsystem:abbreviations}
\begin{align} \label{eq:A}
  \mathbf{A}(t) & \equiv \begin{pmatrix} K_{21} & G_{21} \\ K_{31} & G_{31}  \end{pmatrix},
\end{align}
the inhomogeneity
\begin{align} \label{eq:B}
  \mathbf{B}(t) & \equiv \begin{pmatrix} K_{22} & K_{23} & G_{22} & G_{23} \\ K_{32} & K_{33} & G_{32} & G_{33} \end{pmatrix},
\end{align}
the initial value vector
\begin{align} \label{eq:J}
  \hat{\mathbf{J}} & \equiv ( \hat{X}_1(0), \hat{X}_2(0), \hat{P}_1(0), \hat{P}_2(0) )^{T},
\end{align}
\end{subequations}
and the noises $\hat{\Lambda}_1(t)$ and $\hat{\Lambda}_2(t)$, respectively, from Eq.~(\ref{eq:noise}). In Eqs.~(\ref{eq:A}) and (\ref{eq:B}), $G_{kl}(t)$ and $K_{kl}(t)$ represent the elements in the $k$th row and $l$th column of the respective propagator matrices $\mathbf{G}(t)$ and $\mathbf{K}(t)$, Eq.~(\ref{eq:propagators}). To simplify our notation, recurring time dependencies of these matrix elements have been omitted.\par
In particular, Eq.~(\ref{eq:solutions:PX}) highlights the key aspect of pointer-based simultaneous measurements: the pointer positions $\hat{X}_1(t)$ and $\hat{X}_2(t)$ contain information on the initial system observables $\hat{X}_S(0)$ and $\hat{P}_S(0)$. This structural behavior builds the framework for the following section, where we present an appropriate way to retrieve the information about the initial system observables from the independent pointer positions and discuss the associated measurement uncertainty and its boundaries.


\section{Uncertainty of open pointer-based simultaneous measurements} \label{sec:uncertainty relation}
Our previous considerations have revealed the intimate connection between the pointer positions and the initial system observables. In this section, we now focus on the first and second moments of the so-called inferred observables, which represent a suitable linear combination of the pointer positions in such a way that the expectation values of the inferred observables correspond to the expectation values of the initial system observables. As a consequence, reading out the inferred observables corresponds to a measurement of the initial system observables with an uncertainty given by the variance of the inferred observables. From a closer inspection of this uncertainty follows a lower bound, which can be viewed as the lowest possible uncertainty of a noisy measurement \footnote{A general treatment of uncertainty relations for the quantum Brownian motion of multipartite systems can be found in Ref.~\cite{anastopoulos2010}.}.

\subsection{Inferred observables}
To retrieve the initial system observables from the pointer positions, one can rewrite Eq.~(\ref{eq:solutions:PX}) in the form
\begin{align} \label{eq:inferredsystem}
  \begin{pmatrix} \hat{X}_S(0) \\ \hat{P}_S(0) \end{pmatrix} = & \phantom{-.} \begin{pmatrix} \hat{\mathcal{X}}(t) \\ \hat{\mathcal{P}}(t) \end{pmatrix} \nonumber \\
  & - \mathbf{A}^{-1}(t) \mathbf{B}(t) \hat{\mathbf{J}} - \mathbf{A}^{-1}(t) \begin{pmatrix} \hat{\Lambda}_1(t) \\ \hat{\Lambda}_2(t) \end{pmatrix},
\end{align}
where we have introduced a new pair of observables, namely, the so-called inferred position observable $\hat{\mathcal{X}}(t)$ and the inferred momentum observable $\hat{\mathcal{P}}(t)$ with the rescaling
\begin{align} \label{eq:inferredvariables}
  \begin{pmatrix} \hat{\mathcal{X}}(t) \\ \hat{\mathcal{P}}(t) \end{pmatrix} & \equiv \mathbf{A}^{-1}(t) \begin{pmatrix} \hat{X}_{1}(t) \\ \hat{X}_{2}(t) \end{pmatrix}.
\end{align}
Since the pointer's initial positions $\hat{X}_1(0)$ and $\hat{X}_2(0)$ commute by definition, the inferred observables $\hat{\mathcal{X}}(t)$ and $\hat{\mathcal{P}}(t)$ also commute for later times \footnote{If a renormalization strategy (cf. Sec.~\ref{sec:operational view:renormalization}) is performed without care, it can in fact disturb the unitarity of the time evolution and thus the commutativity of the inferred variables. We assume here that a possible disturbance is small enough to be neglected.}. Moreover, the expectation value of Eq.~(\ref{eq:inferredsystem}) reads
\begin{align} \label{eq:firstmoments:inferred}
  \Braket{ \begin{pmatrix} \hat{\mathcal{X}}(t) \\ \hat{\mathcal{P}}(t) \end{pmatrix} } & = \Braket{ \begin{pmatrix} \hat{X}_S(0) \\ \hat{P}_S(0) \end{pmatrix} } + \mathbf{A}^{-1}(t) \mathbf{B}(t) \braket{ \hat{\mathbf{J}} }
\end{align}
with
\begin{align} \label{eq:elambda}
 \braket{\hat{\mathbf{\Lambda}}(t)} = 0
\end{align}
due to Eq.~(\ref{eq:estochasticforce}). Here, the key feature of the inferred observables assumes shape: Their expectation values represent the corresponding initial expectation values of the system observables, shifted by the initial expectation values of the pointers. However, by choosing appropriate initial pointer states which lead to 
\begin{align} \label{eq:eJ}
\braket{ \hat{\mathbf{J}} } = 0
\end{align}
for the initial values, Eq.~(\ref{eq:J}), this shift can be avoided. To simplify our notation, we assume such pointer states in the following.\par
In conclusion, the inferred observables, Eq.~(\ref{eq:inferredvariables}), can be understood as the effectively measured observables from which the system observables can be directly read out \footnote{In fact, the system observables $\hat{X}_S(0)$ and $\hat{P}_S(0)$ can be inferred from the inferred observables, Eq.~(\ref{eq:inferredvariables}), only if $\det \mathbf{A}(t) \neq 0$. For $\eta \equiv 0$, this condition reads $\kappa_1 \kappa_2 t^2 \neq 0$, which is immediately clear from a physical point of view.}. Therefore, also the uncertainty of the measurement has to be based on these observables.

\subsection{Variances}
In the spirit of the previous considerations, we define the inferred position variance
\begin{subequations} \label{eq:variances}
\begin{align}
  \Delta \mathcal{X}^2(t) \equiv \braket{\hat{\mathcal{X}}^2(t)} - \braket{\hat{\mathcal{X}}(t)}^2
\end{align}
and, analogously, the inferred momentum variance
\begin{align}
  \Delta \mathcal{P}^2(t) \equiv \braket{\hat{\mathcal{P}}^2(t)} - \braket{\hat{\mathcal{P}}(t)}^2.
\end{align}
\end{subequations}
Specifically, Eq.~(\ref{eq:variances}) describes the uncertainty of measuring the initial position and momentum of the system, respectively, by means of a pointer-based simultaneous measurement. Using the variance as an uncertainty measure is a common approach but can nevertheless be considered a controversial subject \cite{hilgevoord1990bialynicki2011}. However, it works perfectly well as long as we deal with localized probability distributions and we therefore stick to this method in the course of this paper.\par
A straightforward calculation using Eqs.~(\ref{eq:inferredsystem}) and (\ref{eq:variances}) shows that the inferred variances
\begin{subequations} \label{eq:variances:1}
\begin{align}
  \Delta \mathcal{X}^2(t) & = \Delta X_S^2(0) + \sigma_1^2(t) + \Xi_{1}^2(t)
\end{align}
and
\begin{align}
  \Delta \mathcal{P}^2(t) & = \Delta P_S^2(0) + \sigma_2^2(t) + \Xi_{2}^2(t)
\end{align}
\end{subequations}
consist of three parts: First, the initial system variances $\Delta X_S^2(0) $, Eq.~(\ref{eq:systemvariance:x}), and $\Delta P_S^2(0) $, Eq.~(\ref{eq:systemvariance:p}), second, the two pointer-based contributions
\begin{align} \label{eq:Delta}
  \sigma_k^2(t) & \equiv \mathbf{v}_k(t) \braket{ \hat{\mathbf{J}} \hat{\mathbf{J}}^{T} } \mathbf{v}_k^{T}(t)
\end{align}
with $k \in \{1,2\}$ and the abbreviation
\begin{align} \label{eq:v}
  \begin{pmatrix} \mathbf{v_1}(t) \\ \mathbf{v_2}(t) \end{pmatrix} \equiv \mathbf{A}^{-1}(t)\mathbf{B}(t),
\end{align}
and third, the covariance matrix of the noise
\begin{align} \label{eq:Xi}
  \mathbf{\Xi}^2(t) & \equiv \mathbf{A}^{-1}(t) \Braket{\begin{pmatrix} \hat{\Lambda}_1 \hat{\Lambda}_1 & \hat{\Lambda}_1 \hat{\Lambda}_2 \\ \hat{\Lambda}_2 \hat{\Lambda}_1 & \hat{\Lambda}_2 \hat{\Lambda}_2 \end{pmatrix}} \mathbf{A}^{-T}(t).
\end{align}
To simplify our notation, we omit recurring time dependencies of the noises $\hat{\Lambda}_1(t)$ and $\hat{\Lambda}_2(t)$, Eq.~(\ref{eq:noise}), in Eq.~(\ref{eq:Xi}). Furthermore, we make use of Eqs.~(\ref{eq:elambda}) and (\ref{eq:eJ}). In Eq.~(\ref{eq:variances:1}), $\Xi^2_{k}(t)$ represents the element in the $k$th row and $k$th column of the covariance matrix of the noise $\mathbf{\Xi}^2(t)$, Eq.~(\ref{eq:Xi}).\par 
In particular, Eq.~(\ref{eq:variances:1}) allows us to directly recognize the separation of the ``intrinsic" or ``preparation uncertainty," namely, $\Delta X_S^2(0) $ and $\Delta P_S^2(0) $, from the ``extrinsic" or ``measurement uncertainty," which can be further diversified into the uncertainty from the (damped) measurement instruments, Eq.~(\ref{eq:Delta}), and the uncertainty from the environmental noise, Eq.~(\ref{eq:Xi}). For a closed pointer-based simultaneous measurement, this separation is well known; cf., e.\,g., Ref.~\cite{appleby1998} and references therein. In our model, we can naturally incorporate the environmental noise as a supplementary extrinsic uncertainty component into the known descriptions.

\subsection{Uncertainty}
The collective uncertainty
\begin{align} \label{eq:U}
  \mathcal{U}^2(t) & \equiv \Delta \mathcal{X}^2(t) \Delta \mathcal{P}^2(t)
\end{align}
is based on the variances of the inferred variables, Eq.~(\ref{eq:variances:1}), and describes the uncertainty of measuring the system's position and the system's momentum by means of a pointer-based simultaneous measurement. Since the collective uncertainty, Eq.~(\ref{eq:U}), is time dependent, there exists at least one optimal measurement time. It is, however, very difficult to analytically optimize $\mathcal{U}^2(t)$ with respect to the measurement time $t$ and we therefore do not further pursue this approach. Instead, we will focus on a formal lower bound which brings out the basic physics of this product of uncertainties and delay the discussion of time dependencies to Sec.~\ref{sec:uncertainty relation:numerics}, where we perform a numerical evaluation.\par
In the case of a closed pointer-based measurement (i.\,e., $\eta = 0$), there exists a constant lower bound of the collective uncertainty, Eq.~(\ref{eq:U}), which reads
\begin{align} \label{eq:U:closed}
  \mathcal{U}^2(t) \geq 1
\end{align}
and represents the combined intrinsic and measurement uncertainty of system and pointers. This result is well known and has been derived with various approaches; see, e.\,g., Refs.~\cite{arthurs1965,she1966,wodkiewicz1987,arthurs1988,ozawa2004,busshardt2010}. Moreover, it fits within the more general framework of the recently found error-disturbance relation \cite{busch2013a} for which it serves as a specific example \cite{busch2007}.\par
At this point one might ask: Does a similar lower uncertainty bound also exist in case of an \emph{open} pointer-based measurement? In order to answer this question, we first use Eq.~(\ref{eq:variances:1}) to rewrite Eq.~(\ref{eq:U}) as
\begin{widetext}
\begin{align}\label{eq:U:1}
  \mathcal{U}^2(t) = & \phantom{+.} \left[ \Delta X_S(0) \sigma_2(t) - \Delta P_S(0) \sigma_1(t) \right]^2 + \frac{\Xi_{2}^2(t)}{2} \Bigg\{ \left[ \Delta X_S(0) + \sigma_1(t) \right]^2 + \left[ \Delta X_S(0) - \sigma_1(t) \right]^2 \Bigg\} + \Xi_{1}^2(t) \Xi_{2}^2(t) \nonumber \\
  & + \left[ \Delta X_S(0) \Delta P_S(0) + \sigma_1(t) \sigma_2(t) \right]^2 + \frac{\Xi_{1}^2(t)}{2} \Bigg\{ \left[ \Delta P_S(0) + \sigma_2(t) \right]^2 + \left[ \Delta P_S(0) - \sigma_2(t) \right]^2 \Bigg\}.
\end{align}
\end{widetext}
In particular, all terms on the right-hand side of Eq.~(\ref{eq:U:1}) are non-negative and we can therefore estimate a lower bound by minimizing the individual terms.\par
The first term, the second part of the second term, and the second part of the fifth term on the right-hand side of Eq.~(\ref{eq:U:1}) vanish if
\begin{subequations}
\begin{align} \label{eq:Umin:1}
  \sigma_1(t) & \overset{!}{=} \Delta X_S(0)
\end{align}
and
\begin{align} \label{eq:Umin:2}
  \sigma_2(t) & \overset{!}{=} \Delta P_S(0)
\end{align}
\end{subequations}
hold true. Additionally, we can utilize Heisenberg's uncertainty relation, which reads
\begin{subequations}
\begin{align} \label{eq:HeisenbergUR:1}
  \Delta X_S(0) \Delta P_S(0) & \geq \frac{1}{2}
\end{align}
for the initial system variances, Eq.~(\ref{eq:systemvariance}), and \footnote{The fundamental lower bound for the pointer-based uncertainty contributions of an open measurement, Eq.~(\ref{eq:HeisenbergUR:2}), is the same as for a closed measurement, cf. Ref.~\cite{busshardt2010}, and can be derived analogously.}
\begin{align} \label{eq:HeisenbergUR:2}
  \sigma_1(t) \sigma_2(t) & \geq \frac{1}{2}
\end{align}
\end{subequations}
for the pointer-based contributions, Eq.~(\ref{eq:Delta}). As a result, we find the lower bound
\begin{align} \label{eq:U:UR}
  \mathcal{U}^2(t) \geq & \phantom{+.} 1 + \Xi_{1}^2(t) \Xi_{2}^2(t) \nonumber \\
  & + \frac{\Xi_{2}^2(t)}{2}[ \Delta X_S(0) + \sigma_1(t) ]^2 \nonumber \\
  & + \frac{\Xi_{1}^2(t)}{2}[ \Delta P_S(0) + \sigma_2(t) ]^2
\end{align}of the collective uncertainty, Eq.~(\ref{eq:U}). It contains the intrinsic and measurement uncertainty from Eq.~(\ref{eq:U:closed}) as well as additional environmental noises. More specifically, the second term represents the pure background noise of the bath, Eq.~(\ref{eq:Xi}), whereas the third and fourth terms can be understood as amplified noises, which are controlled by the initial system variances, Eq.~(\ref{eq:systemvariance}), and the pointer-based variance contributions, Eq.~(\ref{eq:Delta}).\par
We emphasize that the initial system variances are completely independent of the pointer-based variance contributions and the noises, respectively. Thus, the specific structure of this lower bound allows us to determine a distinction within the class of intrinsic minimal uncertainty states. Depending on the values of the pointer-based variance contributions, either coherent states \cite{barnett1997} with $\Delta X_S(0) = \Delta P_S(0)$ or squeezed states \cite{barnett1997} with $\Delta X_S(0) > \Delta P_S(0)$ or $\Delta X_S(0) < \Delta P_S(0)$ will lead to a smaller noise amplification. Hence, depending on the values of the noises, the squeezing of the initial system state determines the minimal collective uncertainty of the measurement. This cannot be seen by looking at the original Heisenberg relation for intrinsic uncertainties nor is it covered by the collective uncertainty relation of a closed simultaneous measurement, Eq.~(\ref{eq:U:closed}).\par
Since the right-hand side of Eq.~(\ref{eq:U:UR}) is in contrast to the right-hand side of Eq.~(\ref{eq:U:closed}) state dependent and therefore strictly speaking not a fundamental bound, it might not be appropriate to call it an uncertainty relation in the sense of Ref.~\cite{busch2013a}. Nevertheless, as a state-dependent disturbance measure \cite{busch2013b}, Eq.~(\ref{eq:U:UR}) can be considered an extension of the uncertainty relation, Eq.~(\ref{eq:U:closed}), which additionally incorporates the dynamically generated disturbance from a noisy measurement device.\par
It is clear that the collective uncertainty reaches the value on the right-hand side of Eq.~(\ref{eq:U:UR}) only if Eqs.~(\ref{eq:Umin:1}) and (\ref{eq:Umin:2}) are fulfilled and equality in Eqs.~(\ref{eq:HeisenbergUR:1}) and (\ref{eq:HeisenbergUR:2}) holds true. However, it is not obvious if one can always find pointer states which fulfill these conditions. Nevertheless, the lower bound, Eq.~(\ref{eq:U:UR}), sets a basic lower limit for the uncertainty in simultaneous measurements and can therefore be considered as the lowest possible uncertainty of a noisy measurement.\par
We furthermore note that in Ref.~\cite{wodkiewicz1987} one can find a lower bound for the uncertainty of a pointer-based measurement which resembles Eq.~(\ref{eq:U:UR}) and is therefore worthwhile to discuss. The author assumes an unspecified measurement device with $n$ degrees of freedom in thermal equilibrium with thermal energy $\beta^{-1}$ and postulates
\begin{align} \label{eq:wodkiewicz:U}
 \mathcal{U}^2 & \geq \frac{1}{4} \left[ 1 + f(n,\beta) \right]^2
\end{align}
with the unspecified function $f(n,\beta)$, ``which depends on the statistical properties of the measurement device." A comparison of Eqs.~(\ref{eq:U:UR}) and (\ref{eq:wodkiewicz:U}) reveals the explicit form
\begin{align} \label{eq:wodkiewicz:f}
  f(n,\beta) = -1 \pm 2 \sqrt{ \mathcal{U}_{\mathrm{min}}^2 },
\end{align}
where $\mathcal{U}_{\mathrm{min}}^2$ represents the right-hand side of Eq.~(\ref{eq:U:UR}). As suggested, Eq.~(\ref{eq:wodkiewicz:f}) indeed depends on the number of oscillators $N \sim n$, which we declared as an infinite continuum in Sec.~\ref{sec:operational view}, and the thermal energy $\beta^{-1}$, Eqs.~(\ref{eq:autocorrelation}), (\ref{eq:noise}), and (\ref{eq:Xi}). Thus, Eq.~(\ref{eq:U:UR}) allows us to clarify the hypothesis (up to this point unproven), Eq.~(\ref{eq:wodkiewicz:U}).

\subsection{Evaluation of the uncertainty of a specific measurement configuration} \label{sec:uncertainty relation:numerics}
\begin{figure}[ht]
  \centering \includegraphics{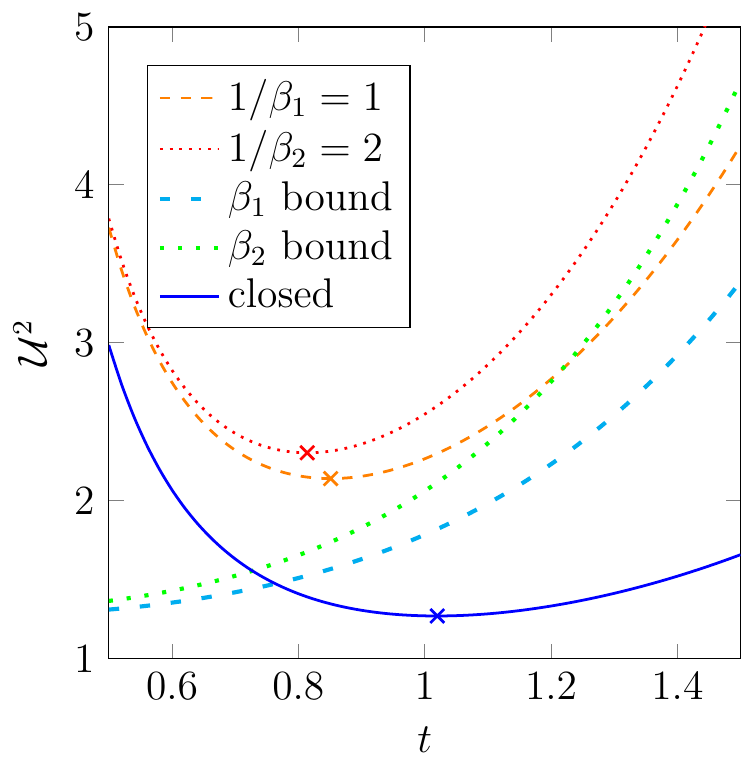}
  \caption{Collective uncertainy ${\mathcal{U}}^2$, Eq.~(\ref{eq:U}), of the measurement configuration discussed in Sec.~\ref{sec:operational view} and its lower bound, Eq.~(\ref{eq:U:UR}), for two different baths of thermal energies $\beta_1^{-1} = 1$ and $\beta_2^{-1} = 2$ as a function of the interaction time $t$. We use initially uncorrelated Gaussian states for both system and pointers as given by Eq.~(\ref{eq:GaussianStates}) and the parameters $\kappa_1 = \kappa_2 = 2$, Eq.~(\ref{eq:H:rescaled}), $M_0 = 1$, Eq.~(\ref{eq:M0}), $\omega_c = 20$ and $\eta = 0.25$, Eq.~(\ref{eq:spectraldensity:ohmic}). At $t\approx1$, the lower bound of the collective uncertainty is closest to the collective uncertainty, while for earlier or later times the distance increases. Markers indicate the optimal measurement time $t_{\mathrm{opt}}$ with $\operatorname{min} {\mathcal{U}}^2(t) = {\mathcal{U}}^2(t_{\mathrm{opt}})$. For higher thermal energies $t_{\mathrm{opt}}$ becomes smaller; cf. Fig.~\ref{fig:numericsresult2}. For the sake of completeness, the collective uncertainty of a closed measurement without any environment (i.\,e., $\eta = 0$) is also shown. Note that this collective uncertainty obeys the constant lower bound given by Eq.~(\ref{eq:U:closed}).}
  \label{fig:numericsresult1}
\end{figure}
\begin{figure}[ht]
  \centering \includegraphics{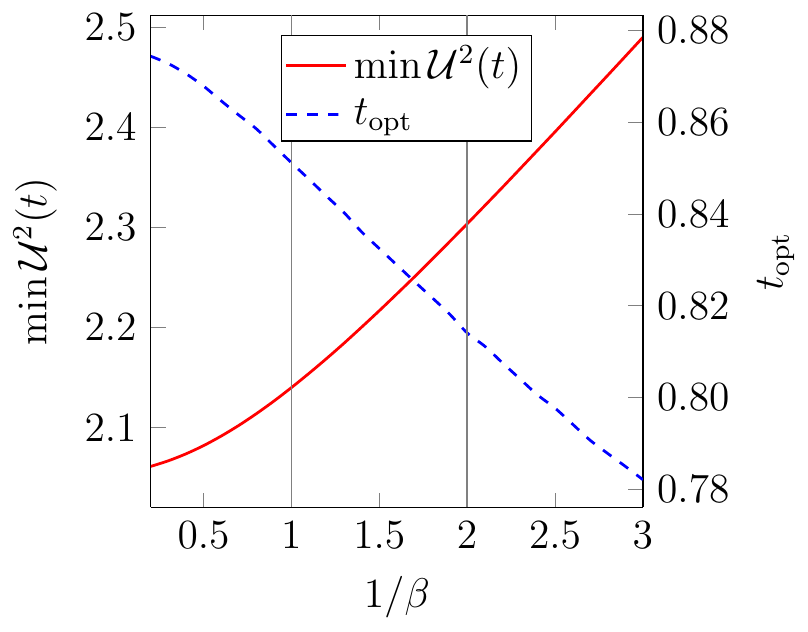}
  \caption{Minimal collective uncertainty $ \operatorname{min} {\mathcal{U}}^2(t) = {\mathcal{U}}^2(t_{\mathrm{opt}})$, Eq.~(\ref{eq:U}), and optimal measurement times $t_{\mathrm{opt}}$ for different thermal energies $\beta^{-1}$. The same configuration as in Fig.~\ref{fig:numericsresult1} is being used. Vertical lines highlight the thermal energies $\beta_1^{-1} = 1$ and $\beta_2^{-1} = 2$ of Fig.~\ref{fig:numericsresult1}. The optimal measurement time becomes smaller for higher thermal energies, while its associated minimal collective uncertainty becomes larger. Both the optimal measurement time and its associated minimal collective uncertainty are roughly proportional to the thermal energy for higher thermal energies and slowly converging for lower thermal energies.}
  \label{fig:numericsresult2}
\end{figure}
Exemplarily, the collective uncertainties, Eq.~(\ref{eq:U}), of two specific measurement configurations and their lower bounds, Eq.~(\ref{eq:U:UR}), are shown in Fig.~\ref{fig:numericsresult1} as the result of numerical calculations \footnote{For practical reasons our numerical calculations are based on a matrix approach instead of a Laplace transform to solve our equations of motion, Eq.~(\ref{eq:X}) and Eq.~(\ref{eq:P}), cf. Sec.~\ref{sec:operational view:formal solution} and Ref.~\cite{fleming2011b,coffey1996}. The specific form of the dissipation kernel, Eq.~(\ref{eq:dissipationkernel}), allows us to express the convolution term in Eq.~(\ref{eq:X}) as the solution of a system of three first-order ordinary differential equations with an inhomogeneity given by the system and pointer velocities. By inserting these differential equations into our original equations of motion and thereby expanding them, we end up with a system of nine inhomogeneous linear ordinary differential equations with constant coefficients, which can be solved straightforwardly with matrix exponential functions.}. For system and pointers we use initially uncorrelated squeezed states \cite{barnett1997} with unity position variance, which can be written as
\begin{align}\label{eq:GaussianStates}
  \braket{x | \psi} = \braket{x | \phi}_1 = \braket{x | \phi}_2 = \left(\frac{e^{-x^2}}{2 \pi}\right)^{\frac{1}{4}}
\end{align}
in position space and fulfill equality in Eq.~(\ref{eq:HeisenbergUR:1}). All other parameters of the measurement configuration are given in the caption of Fig.~\ref{fig:numericsresult1}.\par
From a qualitative point of view, it takes some time for the pointers to accumulate information about the system to be measured, so the collective uncertainty should start high and should begin to shrink over time. On the other hand, if the interaction is too long, the pointer's influence on each other and the bath disturb the measurement results, thus the collective uncertainty should rise again. Therefore, the time-dependent collective uncertainty is expected to have a single minimum value. These considerations are validated by our numerics shown in Fig.~\ref{fig:numericsresult1}.\par
The interaction time at which the collective uncertainty attains its lowest value represents the optimal measurement time $t_{\mathrm{opt}}$ at which the pointers can be read out. In particular, the optimal measurement time $t_{\mathrm{opt}}$ can be treated as the typical measurement time scale of our measurement device $\tau \approx t_{\mathrm{opt}} \approx 1$, Eq.~(\ref{eq:highcutofflimit}), so that $\omega_c \tau \gg 1$ holds true and the high-cutoff-limit approximation, Eq.~(\ref{eq:highcutoff}), is justified.\par
Our results show that at such ``intermediate" interaction times $t\approx1$, the lower bound of the collective uncertainty, Eq.~(\ref{eq:U:UR}), is closest to the collective uncertainty, while for earlier or later times the distance increases. Furthermore, a higher thermal energy $\beta^{-1}$ leads to a smaller optimal measurement time $t_{\mathrm{opt}}$. This is a physically reasonable behavior since the noisy influence of the bath is stronger for higher thermal energies and will increasingly disturb the measurement the longer the interaction between bath and pointers takes place.\par
We strengthen this assumption with our results from Fig.~\ref{fig:numericsresult2}, where we depict the optimal measurement times and their associated minimal collective uncertainty for different thermal energies. As one can see, the optimal measurement time becomes smaller for higher thermal energies, while its associated minimal collective uncertainty becomes larger. Both the optimal measurement time and its associated minimal collective uncertainty are roughly proportional to the thermal energy for higher thermal energies and slowly converging for lower thermal energies.\par
Summarized, the collective uncertainty, Eq.~(\ref{eq:U}), which describes the variance-based uncertainty of a simultaneous pointer-based measurement, is bounded from below by Eq.~(\ref{eq:U:UR}). We can connect this lower bound to the familiar bound of closed pointer-based measurements, Eq.~(\ref{eq:U:closed}), and a bound from Ref.~\cite{wodkiewicz1987}, Eq.~(\ref{eq:wodkiewicz:U}). In particular, Eq.~(\ref{eq:U:UR}) is one of the main results of this paper. Its numerical evaluation reveals that it is best suited for intermediate measurement times; cf. Fig.~\ref{fig:numericsresult1}.


\section{Conclusion} \label{sec:conclusion}
Our model describes a pointer-based simultaneous measurement under the influence of an Ohmic environment with bilinear coupling between all participating particles. The associated equations of motion can be solved formally and reveal a close connection between the pointer positions and the initial system observables. From these solutions, the inferred observables, whose expectation values correspond to the expectation values of the initial system observables and which therefore represent effectively measurable observables arise naturally. Their combined variances determine the collective uncertainty of the simultaneous measurement procedure. Starting from this collective uncertainty, we establish a lower bound for the uncertainty of the so defined noisy measurement. This lower bound is an extension of a previously known uncertainty relation for closed pointer-based measurements and it includes the classic Heisenberg inequality for purely intrinsic uncertainties. Finally, our exemplary evaluation shows that there are optimal measurement times at which the collective uncertainty is minimal.\par
The most restrictive assumption of our model is our postulated Hamiltonian, which includes only terms of quadratic order. Terms of higher order would lead to nonlinear differential equations for which our considerations might not be strictly valid anymore. However, we assume that in a realistic modeling, such terms of higher order will lead only to correction terms of smaller magnitude than the quadratic terms. Therefore, we expect that our model is able to describe the key features of a noisy simultaneous measurement process.\par
Despite the fact that we limited our discussion to the measurement of pointer positions, our model can in principle also be extended to describe the measurement of pointer momenta or any other pair of continuous pointer observables as long as the commutator of these observables does not depend on these observables. Otherwise, nonlinear equations emerge which might lead to different results.\par
Furthermore, we could use our model as a point of origin to discuss different measures of uncertainty for the inferred variables; see, e.\,g., Ref.~\cite{busch2007} and references therein. In particular, using information entropy as an uncertainty measure allows a comparison with previous results for closed pointer-based measurements \cite{heese2013}. Additionally, we could analyze the optimality of different pointer configurations with respect to the measurement uncertainty while taking into account their preparation energy and entanglement \cite{busshardt2011}. Another promising research approach in this context, which we have not yet examined in more detail, is a comparison of the different time scales of our model for decoherence, noise, and system-to-pointer information transfer. Finally, the role of non-Markovian effects in the dynamics has not been explored yet and it might be possible to considerably simplify the equations of motion in certain regimes with the help of a Markovian approximation, cf. Refs.~\cite{fleming2011a,breuer2009} and references therein.\par


%


\end{document}